\documentclass[footinbib,aps,prb,twocolumn,superscriptaddress,groupedaddress]{revtex4}
\usepackage{graphicx}
\usepackage{dcolumn}
\usepackage{bm}
\usepackage{amssymb}
\usepackage{amsmath}
\usepackage{subfigure}
\usepackage{color}
\usepackage{xcolor}
\usepackage{lscape}
\begin{document}


\title{First-principles theory of direct-gap optical emission in hexagonal Ge\\and its enhancement via strain engineering}


\author{Christopher A.~Broderick}
\email{christopher.broderick@ucc.ie} 
\affiliation{Materials Department, University of California, Santa Barbara, California 93106-5050, U.S.A.}
\affiliation{School of Physics, University College Cork, Cork T12 YN60, Ireland}
\affiliation{Tyndall National Institute, University College Cork, Lee Maltings, Dyke Parade, Cork T12 R5CP, Ireland}

\author{Xie Zhang}
\affiliation{School of Materials Science and Engineering, Northwestern Polytechnical University, Xi'an 710072, China}

\author{Mark E.~Turiansky}
\affiliation{Materials Department, University of California, Santa Barbara, California 93106-5050, U.S.A.}

\author{Chris G.~Van de Walle}
\affiliation{Materials Department, University of California, Santa Barbara, California 93106-5050, U.S.A.}

\date{\today}



\begin{abstract}
    The emergence of hexagonal Ge (2H-Ge) as a candidate direct-gap group-IV semiconductor for Si photonics mandates rigorous understanding of its optoelectronic properties. Theoretical predictions of a ``pseudo-direct'' band gap, characterized by weak oscillator strength, contrast with a claimed high radiative recombination coefficient $B$ comparable to conventional (cubic) InAs. We compute $B$ in 2H-Ge from first principles and quantify its dependence on temperature, carrier density and strain. For unstrained 2H-Ge, our calculated spontaneous emission spectra corroborate that measured photoluminescence corresponds to direct-gap emission, but with $B$ being approximately three orders of magnitude lower than in InAs. We confirm a pseudo-direct- to direct-gap transition under $\sim 2$\% [0001] uniaxial tension, which can enhance $B$ by up to three orders of magnitude, making it comparable to that of InAs. Beyond quantifying strong enhancement of $B$ via strain engineering, our analysis suggests the dominance of additional, as-yet unquantified recombination mechanisms in this nascent material.
\end{abstract}

\maketitle


\section*{Introduction}

Si photonics serves as an enabling platform for applications ranging from datacoms and optical computing to sensing and quantum computing. \cite{Zhou_eLight_2023} However, despite significant progress, the indirect band gaps of Si and Ge render them inefficient light emitters, limiting the realization of active photonic components including light-emitting diodes (LEDs), laser and optical interconnects for monolithic integration on Si. It has therefore remained a persistent objective to develop novel direct-gap group-IV semiconductors compatible with complementary metal-oxide semiconductor (CMOS) fabrication. \cite{Geiger_FM_2015}

Advances in nanowire (NW) fabrication have enabled growth of Ge in the metastable lonsdaleite (2H, ``hexagonal diamond'') phase. \cite{Hauge_NL_2017} Predictions of a ``pseudo-direct'' fundamental band gap in 2H-Ge \cite{De_JPCM_2014,Rodl_PRM_2019} -- originating from back-folding of the L$_{6c}$ conduction band (CB) minimum of conventional cubic (3C, diamond-structured) Ge and characterized by weak oscillator strength \cite{Rodl_PRM_2019} -- have been confirmed via experimental demonstrations of direct-gap photoluminescence \cite{Fadaly_Nature_2020} (PL) and stimulated emission \cite{vanTilburg_CP_2024} from 2H-Ge NWs. Exploiting crystal structure as a novel degree of freedom for band-structure engineering to realize a CMOS-compatible direct-gap semiconductor is driving a surge of interest in 2H-Ge as a light-emitter for Si photonics. It is therefore critical to develop a detailed understanding of radiative recombination in 2H-Ge, to interpret experimental measurements and inform development of this nascent material.

In this Article, we analyze radiative recombination in 2H-Ge via first-principles calculations. We compute the spontaneous emission (SE) rate, which we compare to measured PL spectra. \cite{Fadaly_Nature_2020} Via the integrated SE rate we explicitly compute the radiative recombination coefficient $B$ and lifetime $\tau_{\scalebox{0.7}{\text{rad}}}$, and quantify their dependence on carrier density $n$ and temperature $T$. We also investigate the impact of [0001] uniaxial tension on $B$ and $\tau_{\scalebox{0.7}{\text{rad}}}$, confirming that a strain-induced pseudo-direct- to direct-gap transition \cite{Suckert_PRM_2021} drives significant enhancement of $B$. The SE calculations corroborate that measured PL from 2H-Ge NWs corresponds to radiative recombination across the fundamental (direct) band gap. However, our analysis suggests that the initial assumption of purely radiative recombination in 2H-Ge, which underpinned inference of a high $B$ coefficient, is not consistent with theory. Calculated $\tau_{\scalebox{0.7}{\text{rad}}}$ values suggest that the measured carrier lifetime is dominated by as-yet unquantified recombination mechanisms. Application of uniaxial tension along the [0001] NW axis can enhance $B$ by up to three orders of magnitude, making it comparable to that in a conventional (cubic) direct-gap III-V semiconductor.


\section*{Unstrained 2H-Ge}

We employ the band structure and momentum matrix elements calculated using density functional theory (DFT) -- via the Tran-Blaha modified Becke-Johnson (TB-mBJ) exchange potential -- to compute SE spectra, with the $B$ coefficient then computed via the integrated SE rate (cf.~Methods).


\begin{figure*}[t!]
	\includegraphics[width=1.00\textwidth]{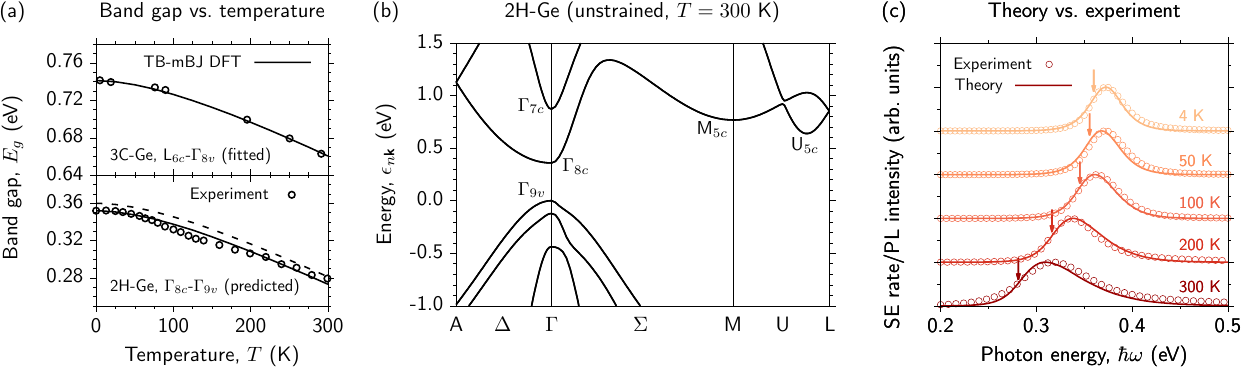}
	\caption{(a) Band gap vs.~temperature for cubic 3C-Ge (upper) and hexagonal 2H-Ge (lower), from parameterized TB-mBJ DFT calculations (lines) and experiment (circles). The dashed line is the calculated 2H-Ge band gap; the solid line has been redshifted to align with experiment at $T = 0$ K. Experimental data for 3C-Ge and 2H-Ge are from Refs.~\onlinecite{Thurmond_JES_1975} and~\onlinecite{Fadaly_Nature_2020}, respectively. (b) DFT-calculated band structure of unstrained 2H-Ge, with TB-mBJ parameterization corresponding to $T = 300$ K. (c) Measured PL (circles) and calculated SE (lines) spectra for unstrained 2H-Ge, for temperatures $T =$ 4, 50, 100, 200 and 300 K. Vertical arrows denote the calculated band-gap energy at each temperature (as in (a)). Experimental PL data are from Ref.~\onlinecite{Fadaly_Nature_2020}.}
	\label{fig:unstrained_2H-Ge_benchmark}
\end{figure*}

In a direct-gap semiconductor, $B \propto \widetilde{f}_{\Gamma} \, E_{g}^{2} \, T^{-D/2}$ at temperature $T$ in $D$ spatial dimensions, \cite{Landsberg_book_1991} where $E_{g}$ is the band gap and $\widetilde{f}_{\Gamma}$ is the polarization-averaged zone-center oscillator strength between the CB and VB edge states. The 2H-Ge band gap has been measured to decrease from 0.352 eV at $T = 0$ K to 0.279 eV at $T = 300$ K. \cite{Fadaly_Nature_2020} Since $B \propto E_{g}^{2}$, given this sizeable $\approx 21$\% decrease of the 2H-Ge band gap up to room temperature, we include a $T$-dependent band gap in our DFT calculations. We achieve this by treating the Becke-Roussel mixing parameter $c_{\scalebox{0.7}{\text{BR}}}$ in the TB-mBJ exchange potential as an empirical parameter, which we fit to the experimental $T$-dependent fundamental (indirect) L$_{6c}$-$\Gamma_{8v}$ band gap of conventional cubic 3C-Ge. \cite{Thurmond_JES_1975} Keeping this fit $c_{\scalebox{0.7}{\text{BR}}} (T)$ fixed, we then apply it to predict the $T$-dependent band gap of 2H-Ge. The results of these calculations are shown in Fig.~\ref{fig:unstrained_2H-Ge_benchmark}(a). Our calculated $T = 0$ K ($c_{\scalebox{0.7}{\text{BR}}} = 1.215$) 2H-Ge band gap of 0.360 eV is in excellent quantitative agreement with experiment, exceeding the measured band gap by only 8 meV. This simple treatment implicitly encapsulates the lattice thermal expansion and electron-phonon coupling that drive the reduction of $E_{g}$ with increasing $T$, accurately capturing the measured $T$-dependent band gap of 2H-Ge. Figure~\ref{fig:unstrained_2H-Ge_benchmark}(b) shows the calculated band structure of 2H-Ge, with $c_{\scalebox{0.7}{\text{BR}}}$ ($= 1.185$) chosen to correspond to $T = 300$ K, where our calculated $E_{g} = 0.281$ eV exceeds the measured value by only 2 meV. We note the presence of a direct fundamental band gap, between the $\Gamma_{8c}$ CB and $\Gamma_{9v}$ VB extrema, with the former originating from back-folding of the L$_{6c}$ CB minimum of 3C-Ge (visible in the lowest CB along $\Delta$ in Fig.~\ref{fig:unstrained_2H-Ge_benchmark}(b)). \cite{De_JPCM_2014,Rodl_PRM_2019}

To validate our analysis, we compare calculated SE spectra $r_{\scalebox{0.7}{\text{sp}}} ( \hbar \omega )$ for unstrained 2H-Ge to the PL measurements of Ref.~\onlinecite{Fadaly_Nature_2020}. The NWs of Ref.~\onlinecite{Fadaly_Nature_2020} have diameters in excess of 150 nm, such that no quantum confinement effects are expected to impact the optoelectronic properties, thereby allowing to compare our bulk DFT calculations to experiment. The results of this qualitative comparison are summarized in Fig.~\ref{fig:unstrained_2H-Ge_benchmark}(c), for $T = 4$ -- 300 K. Beginning at $T =$ 4 K we adjust the carrier density $n$ so that the calculated SE peak energy aligns with the measured PL peak energy, yielding $n = 3 \times 10^{17}$ cm$^{-3}$. Next, we adjust the spectral width $\delta$ of the hyperbolic secant lineshape used to compute $r_{\scalebox{0.7}{\text{sp}}} ( \hbar \omega )$ (cf.~Methods), finding that $\delta = 14$ meV provides a good description of the measured PL lineshape. The remaining SE spectra in Fig.~\ref{fig:unstrained_2H-Ge_benchmark}(c) are then computed by increasing $T$ while keeping $n$ and $\delta$ fixed, motivated by the fact that the PL spectra were measured at fixed excitation intensity ($I = 1.9$ kW cm$^{-2}$). We note that the $T$-dependent PL peak energy is distinct from the $E_{g} (T)$ data of Fig.~\ref{fig:unstrained_2H-Ge_benchmark}(a); the latter were extracted via Lasher-Stern-W\"{u}rfel fits to the measured PL spectra. \cite{Fadaly_Nature_2020} Similarly, the SE peak at temperature $T$ is blueshifted with respect to $E_{g} (T)$ (vertical arrows, Fig.~\ref{fig:unstrained_2H-Ge_benchmark}(c)) due to the Burstein-Moss effect. The calculated SE spectra track the measured PL spectra accurately with increasing $T$, capturing the redshift of the PL peak and the emergence of a high-energy tail driven by the increasing carrier temperature.

\begin{figure*}[t!]
	\includegraphics[width=1.00\textwidth]{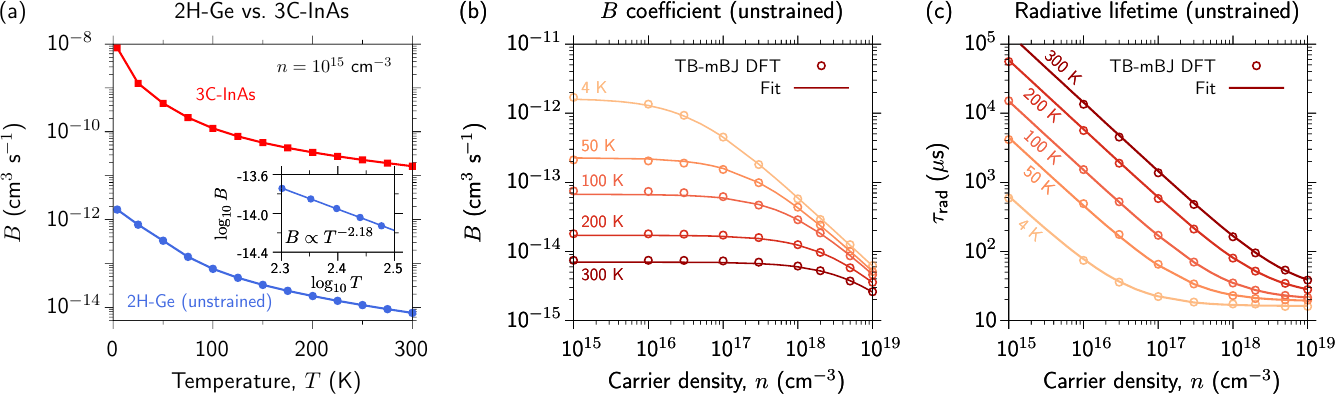}
	\caption{(a) $B$ coefficient vs.~temperature $T$ for unstrained 2H-Ge (blue circles) and 3C-InAs (red squares) at carrier density $n = 10^{15}$ cm$^{-3}$. Inset: log-log plot of $B$ vs.~$T$ in 2H-Ge for $T = 200$ -- 300 K (circles); the line is a linear fit of slope $x$, where $B \propto T^{x}$. (b) $B$ coefficient vs.~$n$ for 2H-Ge at $T = 4$, 50, 100, 200 and 300 K. Circles denote DFT-calculated values of $B$; lines are fits following Eq.~\eqref{eq:radiative_recombination_coefficient_empirical}. (c) Radiative lifetime $\tau_{\scalebox{0.7}{\text{rad}}}$ vs.~$n$ for 2H-Ge at $T = 4$, 50, 100, 200 and 300 K. Circles denote $\tau_{\scalebox{0.7}{\text{rad}}}$ computed using the DFT-calculated $B$ values of (b) via $\tau_{\scalebox{0.7}{\text{rad}}}^{-1} = B n$; lines were obtained similarly via the Eq.~\eqref{eq:radiative_recombination_coefficient_empirical} fits of (b).}
	\label{fig:unstrained_2H-Ge_radiative}
\end{figure*}

Having validated our theoretical approach vs.~experiment for unstrained 2H-Ge, we turn our attention to the $B$ coefficient. Figure~\ref{fig:unstrained_2H-Ge_radiative}(a) compares the calculated $B$ coefficient vs.~$T$ of 3C-InAs and 2H-Ge in the non-degenerate regime (i.e.~at low carrier density, $n = 10^{15}$ cm$^{-3}$). 3C-InAs is chosen as a reference material due to its possessing a narrow, direct band gap ($= 0.354$ eV at $T = 300$ K \cite{Vurgaftman_JAP_2001}) comparable to the pseudo-direct fundamental band gap of 2H-Ge. We note that our calculated $B = 1.64 \times 10^{-11}$ cm$^{3}$ s$^{-1}$ for 3C-InAs at $T = 300$ K is in excellent quantitative agreement with the value $B = 1.81 \times 10^{-11}$ cm$^{3}$ s$^{-1}$ recently reported by Hader et al., \cite{Hader_APL_2022} who employed input from DFT calculations to solve the many-body semiconductor luminescence equations. At $T = 300$ K we calculate $B = 7.44 \times 10^{-15}$ cm$^{3}$ s$^{-1}$ for 2H-Ge, predicting that the room temperature $B$ coefficient of unstrained 2H-Ge is approximately three orders of magnitude lower than that of 3C-InAs.


This low $B$ coefficient is a consequence of the low $\Gamma_{8c}$-$\Gamma_{9v}$ optical (momentum) matrix element, due to the fact that the $\Gamma_{8c}$ CB minimum is a back-folded L$_{6c}$ state from 3C-Ge. This $\Gamma_{8c}$ state thus contains predominantely $p$-like orbital character, closely matching its symmetry to that of the $p$-like $\Gamma_{9v}$ VB maximum. For 2H-Ge at $T = 4$ K we compute oscillator strength $\widetilde{f} ( \Gamma_{8c}\text{-}\Gamma_{9v} ) = E_{P} / E_{g} = 6.81 \times 10^{-3}$, where $E_{P} = \vert \widetilde{p} ( \Gamma_{8c}\text{-}\Gamma_{9v} ) \vert^{2} / 2 m_{0} = 2.45$ meV is the Kane parameter and $\widetilde{p} ( \Gamma_{8c}\text{-}\Gamma_{9v} )$ is the polarization-averaged interband momentum matrix element. This oscillator strength is in close agreement with $\widetilde{f} ( \Gamma_{8c}\text{-}\Gamma_{9v} ) = 3.95 \times 10^{-3}$ obtained by polarization-averaging the TB-mBJ results of Ref.~\onlinecite{Rodl_PRM_2019}. We note that our analysis confirms the optical selection rules identified in Ref.~\onlinecite{Rodl_PRM_2019}, consistent with recent polarization-resolved PL measurements. \cite{vanTilburg_JAP_2023}

To quantify the $T$ dependence of $B$ we assume $B \propto T^{x}$ and compute $x$ as the slope of a linear fit to $\log B$ vs.~$\log T$. This is shown in the inset to Fig.~\ref{fig:unstrained_2H-Ge_radiative}(a) for $T = 200$ -- 300 K, where we compute $B \propto T^{-2.18}$ for 2H-Ge (vs.~$B \propto T^{-1.80}$ for 3C-InAs). This exceeds the $B \propto T^{-3/2}$ dependence expected for an idealized bulk semiconductor, and is partially a consequence of capturing the measured $E_{g} (T)$ in our calculations. Increasing to a degenerate carrier density $n = 10^{18}$ cm$^{-3}$ we calculate $B \propto T^{-1.76 (-1.47)}$ for 2H-Ge (3C-InAs), suggesting that the radiative recombination rate $\tau_{\scalebox{0.7}{\text{rad}}}^{-1} = B n$ in 2H-Ge decreases more rapidly with increasing $T$ than in 3C-InAs.

Figure~\ref{fig:unstrained_2H-Ge_radiative}(b) summarizes the $B$ coefficient vs.~$n$, where open circles denote the results of our DFT calculations. At fixed $T$ we note a strong reduction of $B$ as $n$ is increased beyond the non-degenerate regime, in line with the expected impact of phase-space filling. \cite{Hader_APL_2005,Hader_SPIE_2006} For example, at $T = 300$ K we compute that $B$ decreases weakly between $n = 10^{15}$ and $10^{18}$ cm$^{-3}$, from 7.44 to $6.14 \times 10^{-15}$ cm$^{3}$ s$^{-1}$, beyond which carrier density it decreases rapidly due to phase-space filling. At fixed $T$ this phase-space filling can be parameterized by fitting to the DFT-calculated $B$ vs.~$n$ using the empirical relation \cite{Grein_JAP_2002}

\begin{equation}
    B (n) = \frac{ B_{0} }{ 1 + \frac{ n }{ n_{0} } } \, ,
    \label{eq:radiative_recombination_coefficient_empirical}
\end{equation}

\noindent
where $B_{0}$ is the best-fit value of $B$ in the non-degenerate regime. The results of this fitting are shown using solid lines in Fig.~\ref{fig:unstrained_2H-Ge_radiative}(b) where, e.g., we obtain $B_{0} = 1.63 \times 10^{-12}$ ($7.01 \times 10^{-15}$) cm$^{3}$ s$^{-1}$ and $n_{0} = 3.78 \times 10^{16}$ ($5.80 \times 10^{18}$) cm$^{-3}$ at $T = 4$ (300) K.

Using the DFT-calculated (circles) and Eq.~\eqref{eq:radiative_recombination_coefficient_empirical} fits (lines) of Fig.~\ref{fig:unstrained_2H-Ge_radiative}(b), we compute the radiative lifetime $\tau_{\scalebox{0.7}{\text{rad}}}$ vs.~$n$ at fixed $T$ via $\tau_{\scalebox{0.7}{\text{rad}}}^{-1} = B n$. The resulting lifetimes are summarized in Fig.~\ref{fig:unstrained_2H-Ge_radiative}(c). At $n = 10^{15}$ cm$^{-3}$ we compute $\tau_{\scalebox{0.7}{\text{rad}}} = 0.59$ ms at $T = 4$ K, in good agreement with the values calculated by R\"{o}dl et al. \cite{Rodl_PRM_2019} and Suckert et al. \cite{Suckert_PRM_2021} As $T$ increases, our calculated non-degenerate $\tau_{\scalebox{0.7}{\text{rad}}}$ increases strongly, reaching 135 ms at $T = 300$ K. This diverges strongly from Refs.~\onlinecite{Rodl_PRM_2019} and \onlinecite{Suckert_PRM_2021}, which predicted a $T$-independent $\tau_{\scalebox{0.7}{\text{rad}}}$ up to $T \approx 400$ K. We note that the $T$ independence predicted in Refs.~\onlinecite{Rodl_PRM_2019} and \onlinecite{Suckert_PRM_2021} is a consequence of employing Maxwell-Boltzmann statistics, which artificially pins the quasi-Fermi levels by enforcing unit occupancy at the band extrema -- i.e.~generating an explicit change of $n$ as $T$ is varied. The use of Maxwell-Boltzmann statistics therefore precludes reliable $T$-dependent prediction of $B = ( \tau_{\scalebox{0.7}{\text{rad}}} \, n )^{-1}$. This contrasts with the use of Fermi-Dirac statistics in the present analysis, which allows to independently vary $n$ and $T$. At $T = 300$ K we expect that $\tau_{\scalebox{0.7}{\text{rad}}}$ exceeds its $T = 4$ K value by a factor $B ( 4~\text{K} )/B ( 300~\text{K} )$, with $\tau_{\scalebox{0.7}{\text{rad}}} \propto T^{+2.18}$ at fixed $n$ in 2H-Ge. The strong increase of $\tau_{\scalebox{0.7}{\text{rad}}}$ with increasing $T$ arises in our analysis due to the decrease (increase) in electron (hole) quasi-Fermi level, relative to the CB minimum (VB maximum), required to maintain $n = 10^{15}$ cm$^{-3}$ with increasing $T$. The calculated $\tau_{\scalebox{0.7}{\text{rad}}}$ at fixed $T$ decreases strongly with increasing $n$, reaching a value of 17.4 $\mu$s (162.8 $\mu$s) at $T = 4$ (300) K for $n = 10^{18}$ cm$^{-3}$.

We note that these calculated radiative lifetimes exceed the measured 2H-Ge carrier lifetime, $\tau = 0.98$ (0.46) ns at $T = 4$ (300) K, \cite{Fadaly_Nature_2020} by approximately four orders of magnitude. We recall that our calculated SE spectra qualitatively, but closely, reproduce the measured $T$-dependent PL spectra of Ref.~\onlinecite{Fadaly_Nature_2020}. This supports the key conclusion of Ref.~\onlinecite{Fadaly_Nature_2020}, that the observed PL is consistent with direct-gap band-to-band optical emission from bulk-like 2H-Ge. However, our analysis -- which is consistent with the low-temperature DFT calculations of Refs.~\onlinecite{Rodl_PRM_2019} and~\onlinecite{Fadaly_Nature_2020}, but which allows to accurately predict the $B$ coefficient and its temperature dependence -- indicates that the measured carrier lifetime $\tau$ is not consistent with the expected radiative lifetime. Firstly, $\tau_{\scalebox{0.7}{\text{rad}}}$ for a direct-gap semiconductor is expected to increase with increasing $T$; the data of Ref.~\onlinecite{Fadaly_Nature_2020} demonstrate an $\approx 50$\% decrease of $\tau$ between $T = 4$ and 300 K. Secondly, a high 2H-Ge $B$ coefficient, comparable to that of direct-gap 3C-InAs or 3C-GaAs, was inferred in Ref.~\onlinecite{Fadaly_Nature_2020} by assuming $\tau = \tau_{\scalebox{0.7}{\text{rad}}}$ (i.e.~by assuming an internal quantum efficiency of 100\%). Our explicit calculation of $B$ for 2H-Ge -- beyond the previously investigated non-degenerate regime \cite{Fadaly_Nature_2020,Suckert_PRM_2021} -- suggests that $B$ is approximately three orders of magnitude lower than inferred in Ref.~\onlinecite{Fadaly_Nature_2020}. The fundamental origin of this behavior is the weak $\Gamma_{8c}$-$\Gamma_{9v}$ optical matrix element. Our calculated $E_{P} = 2.45$ meV is at odds with the recent value $E_{P} \geq 3.8$ eV estimated by van Lange et al. \cite{van_Lange_ACSP_2024} However, we note that the source of this discrepancy is as described above: the analysis of van Lange et al.~is predicated on the same underlying assumption of Ref.~\onlinecite{Fadaly_Nature_2020}, that $\tau = \tau_{\scalebox{0.7}{\text{rad}}}$. The oscillator strength associated with the fundamental direct band gap of 2H-Ge thus remains a topic of active investigation. As an experimental test of these competing interpretations, we note that our calculated SE spectra predict that -- at fixed excitation and temperature -- the PL intensity emitted by a 2H-Ge NW should be approximately three orders of magnitude lower than that emitted by an equivalent 3C- or 2H-InAs NW.

Generally, we expect $\tau^{-1} = \tau_{\scalebox{0.7}{\text{rad}}}^{-1} + \tau_{\scalebox{0.7}{\text{non-rad}}}^{-1}$ for the carrier recombination rate. $\tau_{\scalebox{0.7}{\text{rad}}}^{-1}$ can include contributions from phonon-assisted radiative recombination in addition to the direct ($\Delta \mathbf{k} = 0$) radiative recombination considered herein. We note the close theory-experiment correspondence of Fig.~\ref{fig:unstrained_2H-Ge_benchmark}(c), where the absence of visible phonon replicas in the measured PL spectra indicates that phonon-assisted radiative recombination does not contribute appreciably to $\tau_{\scalebox{0.7}{\text{rad}}}^{-1}$. $\tau_{\scalebox{0.7}{\text{non-rad}}}^{-1}$ is the total non-radiative recombination rate, and can include contributions from Auger-Meitner recombination, defect-related (Shockley-Read-Hall) recombination and/or surface recombination at NW facets. \cite{Willem-Jan_Berghuis_ACSANM_2024} Initial investigations of native defects in 2H-Ge NWs have not confirmed the presence of localized trap states lying within the 2H-Ge band gap. \cite{Fadaly_NL_2021} However, we note the presence of a weak ``s-shape'' in the experimental $T$-dependent 2H-Ge band gap (cf.~Fig.~\ref{fig:unstrained_2H-Ge_benchmark}(a)). This is typically a signature of trapped carriers undergoing thermionic emission from near-band-edge localized states with increasing $T$. \cite{Imhof_APL_2010} The measured $\tau \lesssim 1$ ns carrier lifetime suggests the presence of additional recombination mechanisms that, while evidently not precluding direct-gap optical emission, nonetheless dominate recombination in first-generation 2H-Ge NWs. Further work is therefore required to quantify additional recombination mechanisms in 2H-Ge.


\section*{Strained 2H-Ge}

We now turn our attention to the possibility of increasing $B$ (decreasing $\tau_{\scalebox{0.7}{\text{rad}}}$) via strain engineering. Suckert et al.~\cite{Suckert_PRM_2021} predicted theoretically that application of [0001] uniaxial tension to 2H-Ge shifts the pseudo-direct $\Gamma_{8c}$-$\Gamma_{9v}$ (direct $\Gamma_{7c}$-$\Gamma_{9v}$) band gap upwards (downwards) in energy. This produces a pseudo-direct- to direct-gap transition for tensile $c$-axis strain $\epsilon_{zz}$ close to 2\%, such that the fundamental band gap becomes optically bright, leading to a strong reduction of $\tau_{\scalebox{0.7}{\text{rad}}}$. \cite{Suckert_PRM_2021} We note that, due to the close structural similarity of [111] 3C-Ge and [0001] 2H-Ge, this transition corresponds to the known ability to drive an indirect- to direct-gap transition in cubic 3C-Ge via uniaxial [111] tension. \cite{Zhang_PRL_2009}

\begin{figure}[t!]
	\includegraphics[width=0.95\columnwidth]{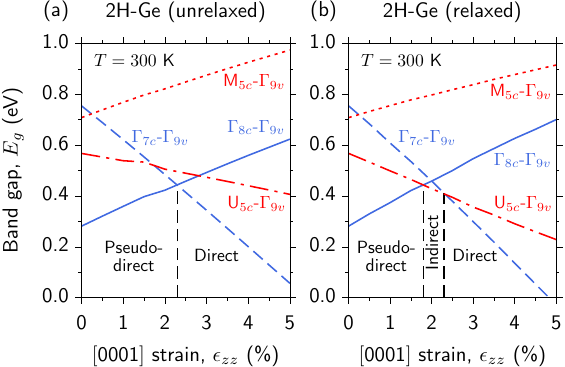}
	\caption{Evolution with [0001] uniaxial tensile strain $\epsilon_{zz}$ of the $T = 300$ K 2H-Ge band gaps between the $\Gamma_{9v}$ VB maximum and the pseudo-direct $\Gamma_{8c}$ (solid blue), direct $\Gamma_{7c}$ (dashed blue), indirect U$_{5c}$ (dash-dotted red), and indirect M$_{5c}$ (dotted red) CB minima. (a) Band gaps vs.~applied [0001] strain $\epsilon_{zz}$, but without allowing relaxation of the $c$-plane lattice parameter $a$ or internal parameter $u$. (b) As in (a), but including relaxation of the lattice and internal parameters $a$ and $u$.}
	\label{fig:2H-Ge_strained_band_gaps}
\end{figure}

Figure~\ref{fig:2H-Ge_strained_band_gaps} shows the calculated band gaps of 2H-Ge under [0001] uniaxial tension for (a) an idealized case in which applying a $c$-axis strain in isolation produces lattice parameter $c = ( 1 + \epsilon_{zz} ) \, c_{0}$, and (b) the physically realistic case in which the uniaxial stress applied to bring about the same extension of $c$ is accompanied by relaxation of the $c$-plane lattice parameter $a$ and internal parameter $u$. Case (b) is equivalent to pseudomorphic growth of compressively strained 2H-Ge on an [0001]-oriented substrate, where lattice relaxation in response to $c$-plane compression produces tension along [0001]. In the unrelaxed case we identify a pseudo-direct- (solid blue) to direct-gap (dashed blue) transition for $\epsilon_{zz} = 1.7$\%, confirming the prediction of Suckert et al. \cite{Suckert_PRM_2021} Comparing Figs.~\ref{fig:2H-Ge_strained_band_gaps}(a) and~\ref{fig:2H-Ge_strained_band_gaps}(b) we observe that lattice relaxation drives a more rapid downward shift in energy of the U$_{5c}$ CB minima (dash-dotted red), leading to the emergence of an indirect-gap regime for a narrow range of tensile strains, $\epsilon_{zz} = 2.1$ -- 2.3\%, coinciding with the pseudo-direct $\Gamma_{8c}$ to direct $\Gamma_{7c}$ crossover at $\Gamma$, beyond which the band gap is direct. This contrasts with Ref.~\onlinecite{Suckert_PRM_2021}, in which an indirect-gap regime was not observed in TB-mBJ DFT calculations employing self-consistent evaluation of $c_{\scalebox{0.7}{\text{BR}}}$. Our test calculations utilizing this approach confirm its underestimation of the experimental band gap of unstrained 2H-Ge by $\approx 15$\%, \cite{Rodl_PRM_2019,Suckert_PRM_2021} and suggests that Fig.~3(c) in Ref.~\onlinecite{Suckert_PRM_2021} may correspond to the unrelaxed case. Our analysis suggests that Brillouin zone- (BZ-) edge U-valley CB states can lie close in energy to the CB minimum in direct-gap tensile strained 2H-Ge. From the perspective of LED/laser operation, engineering the $\Gamma_{8c}$-U$_{5c}$ CB valley splitting then becomes an important consideration: the highly degenerate U valleys present a large density of states, which can drive rapid intervalley scattering of $\Gamma$-point electrons to the BZ edge.


\begin{figure*}[t!]
	\includegraphics[width=1.00\textwidth]{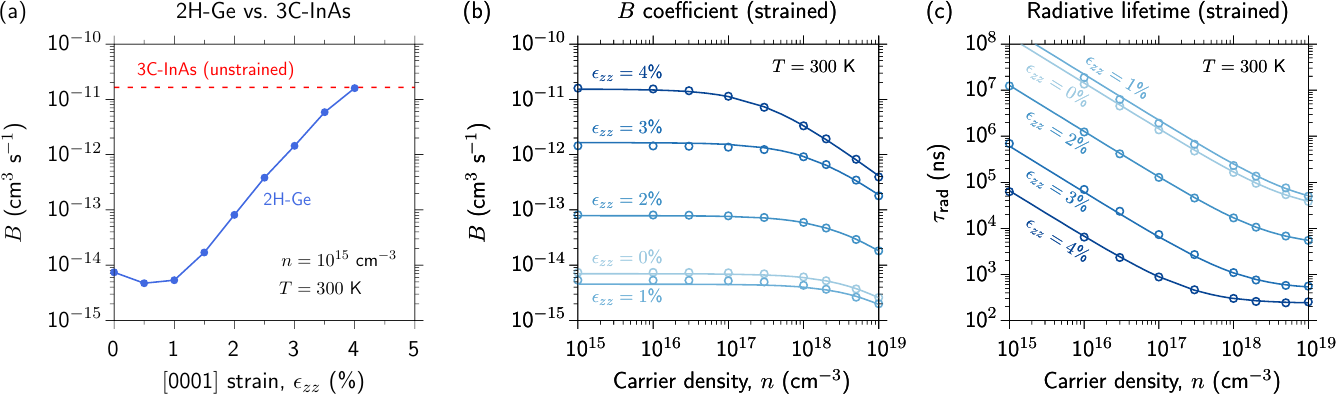}
	\caption{(a) $B$ coefficient vs.~[0001] uniaxial tensile strain $\epsilon_{zz}$ for 2H-Ge at $T = 300$ K and $n = 10^{15}$ cm$^{-3}$ (closed blue circles). The horizontal dashed red line shows the calculated $B$ coefficient of unstrained 3C-InAs at the same $T$ and $n$ (cf.~Fig.~\ref{fig:unstrained_2H-Ge_radiative}(a)). (b) $B$ coefficient vs.~$n$ for 2H-Ge at $\epsilon_{zz} = 0$, 1, 2, 3 and 4\%. Circles denote DFT-calculated values of $B$; lines are fits following Eq.~\eqref{eq:radiative_recombination_coefficient_empirical}. (c) Radiative lifetime $\tau_{\scalebox{0.7}{\text{rad}}}$ vs.~$n$ for 2H-Ge at $\epsilon_{zz} = 0$, 1, 2, 3 and 4\%. Circles denote $\tau_{\scalebox{0.7}{\text{rad}}}$ computed using the DFT-calculated $B$ values of (b) via $\tau_{\scalebox{0.7}{\text{rad}}}^{-1} = B n$; lines were obtained similarly via the Eq.~\eqref{eq:radiative_recombination_coefficient_empirical} fits of (b).}
	\label{fig:strained_2H-Ge_radiative}
\end{figure*}

The results of our strain-dependent analysis are summarized in Fig.~\ref{fig:strained_2H-Ge_radiative}, where we restrict our attention to the relaxed case of Fig.~\ref{fig:2H-Ge_strained_band_gaps}(b). Figure~\ref{fig:strained_2H-Ge_radiative}(a) shows the calculated $B$ coefficient of 2H-Ge vs.~$\epsilon_{zz}$ at $T = 300$ K in the non-degenerate limit (blue). Between $\epsilon_{zz} = 0$ and 1\% we note a decrease in $B$. This is a consequence of the rapid reduction in $\Gamma_{8c}$-U$_{5c}$ splitting (cf.~Fig.~\ref{fig:2H-Ge_strained_band_gaps}(b)), with these BZ-edge states then being partially occupied via the high-energy tail of the electron distribution function. Since we consider only direct radiative transitions, these BZ-edge electrons cannot recombine with zone-center holes, thereby reducing the number of electron-hole pairs available to contribute to the SE rate. As $\epsilon_{zz}$ is increased above 1\% the $\Gamma_{8c}$-$\Gamma_{7c}$ splitting reduces rapidly, leading to an increasing fraction of electrons occupying CB states in the $\Gamma_{7c}$ valley. The $\Gamma_{7c}$-$\Gamma_{9v}$ transition, which derives from the direct $\Gamma_{7c}$-$\Gamma_{8v}$ transition in 3C-Ge, is optically bright. In unstrained 2H-Ge it has $E_{g} = 0.873$ eV and $E_{P} = 18.18$ eV, with oscillator strength $\widetilde{f} ( \Gamma_{7c}\text{-}\Gamma_{9v} ) = 20.82$, which is $\sim 10^{3}$ times larger than that associated with the fundamental (pseudo-direct) $\Gamma_{8c}$-$\Gamma_{9v}$ band gap. For $\epsilon_{zz} > 1$\% recombination via the $\Gamma_{7c}$-$\Gamma_{9v}$ transition dominates the SE rate, leading to a strong increase in $B$. For example, at $\epsilon_{zz} = 3$\% we compute $B = 1.44 \times 10^{-12}$ cm$^{3}$ s$^{-1}$ with direct band gap $E_{g} = 0.299$ eV -- i.e.~band gap close to that of unstrained 2H-Ge, with $B$ increased ($\tau_{\scalebox{0.7}{\text{rad}}}$ decreased) by two orders of magnitude vs.~unstrained 2H-Ge.

Figures~\ref{fig:strained_2H-Ge_radiative}(b) and~\ref{fig:strained_2H-Ge_radiative}(c) respectively show the strain-dependent $B$ and $\tau_{\scalebox{0.7}{\text{rad}}}$ vs.~$n$ at $T = 300$ K. As in Figs.~\ref{fig:unstrained_2H-Ge_radiative}(b) and~\ref{fig:unstrained_2H-Ge_radiative}(c) open circles denotes DFT-calculated values, while solid lines are fits following Eq.~\eqref{eq:radiative_recombination_coefficient_empirical}. As in unstrained 2H-Ge, phase-space filling decreases both $B$ and $\tau_{\scalebox{0.7}{\text{rad}}}$ with increasing $n$. Again selecting $\epsilon_{zz} = 3$\% as an example, as the band gap is close to that of unstrained 2H-Ge, we fit $B_{0} = 1.65 \times 10^{-12}$ cm$^{3}$ s$^{-1}$ and $n_{0} = 1.28 \times 10^{18}$ cm$^{-3}$. Between $n = 10^{15}$ and $10^{18}$ cm$^{-3}$ $B$ decreases weakly, from $1.44 \times 10^{-12}$ to $9.15 \times 10^{-13}$ cm$^{-3}$, beyond which carrier density it decreases rapidly. Similarly, at $\epsilon_{zz} = 3$\% we compute $\tau_{\scalebox{0.7}{\text{rad}}} = 0.69$ ms, decreasing to 1.1 $\mu$s by $n = 10^{18}$ cm$^{-3}$ and 0.56 $\mu$s at $n = 10^{19}$ cm$^{-3}$. We note that these radiative lifetimes are approximately two orders of magnitude lower than those computed for unstrained 2H-Ge above. These results summarize that, in addition to allowing to tune the emission wavelength, strain engineering of 2H-Ge can be expected to strongly enhance $\tau_{\scalebox{0.7}{\text{rad}}}^{-1}$ for mid-infrared emission, provided adequate energy separation between the $\Gamma$- and U-point CB valley minima can be maintained.


\section*{Conclusions}

In summary, we analyzed radiative recombination in lonsdaleite Ge from first principles, including the $B$ coefficient and its dependence on temperature, carrier density and strain. For unstrained 2H-Ge, calculated SE spectra corroborate that experimentally observed PL from 2H-Ge NWs is consistent with direct-gap optical emission. The calculated $B$ coefficient is approximately three orders of magnitude lower than in 3C-InAs, indicating a significant difference between the radiative lifetime $\tau_{\scalebox{0.7}{\text{rad}}}$ and measured carrier lifetime $\tau$, associated with as-yet unquantified recombination mechanisms. We confirmed the presence of a pseudo-direct- to direct-gap transition under [0001] unaxial tension, and elucidated the role played by internal relaxation. We observed an indirect band gap in a narrow strain range straddling this transition, with a direct band gap emerging for tensile strains $\gtrsim 2.3$\%. This highlights CB valley splitting as an important consideration for band-structure engineering of 2H-Ge. We predict that tensile strain strongly enhances $B$, which can approach that of 3C-InAs in the direct-gap regime. Analysis of additional recombination mechanisms is required to inform further development of this emerging semiconductor, with the aim of realizing a direct-gap emitter for Si integrated photonics.


\section*{Methods}
\label{sec:methods}


\paragraph*{Density functional theory:} DFT calculations were performed using the projector-augmented wave (PAW) method, as implemented in the Vienna Ab-initio Simulation Package (\textsc{VASP}). \cite{Blochl_PRB_1994,Kresse_PRB_1999} All calculations use a plane-wave cut-off energy of 500 eV, include spin-orbit coupling, and employ PAW potentials in which the semi-core $(3d)^{10}$ orbitals of Ge and $(4d)^{10}$ orbitals of In are treated as valence states. BZ integration for the 2H and 3C crystal phases was performed using $\Gamma$-centered $11 \times 11 \times 7$ and $11 \times 11 \times 11$ Monkhorst-Pack (MP) $\textbf{k}$-point grids, respectively. The lattice parameters $a_{0} = 3.961$ \AA~and $c_{0} = 6.534$ \AA, and internal parameter $u_{0} = 0.3744$, of 2H-Ge were computed in the LDA and are in good quantitative agreement with previous calculations. \cite{Rodl_PRM_2019} Electronic structure calculations were performed in the meta-generalized gradient approximation, employing the TB-mBJ exchange potential in conjunction with the LDA to the electronic correlation. \cite{Tran_PRL_2009} As described above, the Becke-Roussel mixing parameter $c_{\scalebox{0.7}{\text{BR}}} (T)$ was determined via an empirical fit to the experimental $T$-dependent L$_{6c}$-$\Gamma_{8v}$ band gap of 3C-Ge (cf.~Fig.~\ref{fig:unstrained_2H-Ge_benchmark}(a)). For 3C-InAs we use the LDA-calculated lattice parameter $a_{0} = 6.030$ \AA, and employ an empirical fit $c_{\scalebox{0.7}{\text{BR}}} (T)$ to the experimental $T$-dependent fundamental $\Gamma_{6c}$-$\Gamma_{8v}$ band gap. \cite{Shen_PRB_2019,Vurgaftman_JAP_2001}


\paragraph*{Radiative recombination:} Using the TB-mBJ-calculated band structure and interband momentum matrix elements, \cite{Gajdos_PRB_2006} we compute the SE (rate) spectrum in the quasi-equilibrium approximation as \cite{Chang_IEEEJSQTE_1995}

\begin{eqnarray}
	r_{\scalebox{0.7}{\text{sp}}} ( \hbar \omega ) &=& \frac{ e^{2} n_{r} \omega }{ \pi \epsilon_{0} m_{0}^{2} \hbar c^{3} } \sum_{ n_{c}, n_{v} } \int \frac{ \text{d} \textbf{k} }{ ( 2 \pi )^{3} } \, \vert \widetilde{p}_{n_{c}n_{v}\scalebox{0.7}{\textbf{k}}} \vert^{2} \, f ( \epsilon_{ n_{c} \scalebox{0.7}{\textbf{k}} }, F_{e} ) \nonumber \\
    &\times& ( 1 - f( \epsilon_{ n_{v} \scalebox{0.7}{\textbf{k}} }, F_{h} ) ) \, \delta ( \epsilon_{ n_{c} \scalebox{0.7}{\textbf{k}} } - \epsilon_{ n_{v} \scalebox{0.7}{\textbf{k}} } - \hbar \omega ) \, , \label{eq:spontaneous_emission_spectrum}
\end{eqnarray}

\noindent
where $n_{c(v)}$ indexes the CBs (VBs), $\epsilon_{n_{c(v)} \scalebox{0.7}{\textbf{k}}}$ is the CB (VB) energy at wave vector $\textbf{k}$, $f$ is the Fermi-Dirac distribution function at temperature $T$, $F_{e(h)}$ is the electron (hole) quasi-Fermi level at temperature $T$ and electron (hole) carrier density $n$ ($p$), and $\vert \widetilde{p}_{n_{c}n_{v}\scalebox{0.7}{\textbf{k}}} \vert^{2}$ is the (squared) polarization-averaged momentum matrix element between conduction and valence bands $n_{c}$ and $n_{v}$ at wave vector $\textbf{k}$. The Dirac delta distribution in Eq.~\eqref{eq:spontaneous_emission_spectrum}, which imposes conservation of energy, is replaced by a hyperbolic secant lineshape. \cite{Marko_SR_2016} $B$ is computed via the integrated SE rate \cite{Landsberg_book_1991,Murphy_JPDAP_2024}

\begin{equation}
    B = \frac{ 1 }{ np } \int r_{\scalebox{0.7}{\text{sp}}} ( \hbar \omega ) \, \text{d} ( \hbar \omega ) \, ,
    \label{eq:radiative_recombination_coefficient}
\end{equation}

\noindent
where we impose net charge neutrality, $n = p$.

Accurate evaluation of $F_{e}$, $F_{h}$ and $r_{\scalebox{0.7}{\text{sp}}} ( \hbar \omega )$ mandates dense sampling of BZ regions occupied by electrons and holes; we employ the adaptive $\textbf{k}$-point sampling approach of Ref.~\onlinecite{Zhang_ACSEL_2018}. In unstrained 2H-Ge carriers occupy states in the immediate vicinity of $\Gamma$ only. We replace a $\Gamma$-centred portion of the coarse MP grid employed in the TB-mBJ electronic structure calculations -- extending to 25\% (50\%) of the BZ extent parallel (perpendicular) to the [0001] plane -- by dense MP grids containing up to $55 \times 55 \times 33$ $\textbf{k}$-points. Under [0001] tension, the downward shift of the U valleys can result in an appreciable fraction of injected electrons occupying states at and close to the BZ edge (cf.~Fig.~\ref{fig:2H-Ge_strained_band_gaps}). We augment the adaptive grid employed for unstrained 2H-Ge by also replacing $\textbf{k}$ points -- lying within 50\% of the hexagonal BZ side length of the U direction in the (0001) plane, and along the full extent of the BZ parallel to [0001] -- by the $\textbf{k}$ points lying within that region from $\Gamma$-centred MP grids containing up to $72 \times 72 \times 44$ $\textbf{k}$-points. Using the Kohn-Sham wave functions computed on the coarse MP grid as input, $\epsilon_{{n} \scalebox{0.7}{\textbf{k}}}$ and $\vert \widetilde{p}_{n_{c}n_{v}\scalebox{0.7}{\textbf{k}}} \vert^{2}$ are calculated non-self-consistently on the adaptive grid. The integral in Eq.~\eqref{eq:spontaneous_emission_spectrum} is then evaluated via a weighted sum over the adaptive $\textbf{k}$-point grid. \cite{Zhang_ACSEL_2018}


\section*{Author contributions}

This study was devised by C.A.B., with input from C.G.VdW.~and X.Z. All calculations were performed by C.A.B, with technical support from X.Z.~and M.E.T., and supervised by C.G.VdW. The writing of the manuscript was led by C.A.B., with contributions from all authors.


\section*{Acknowledgements}

C.A.B.~was supported by the European Commission's Horizon 2020 research and innovation program via a Marie Sk\l{}odowska-Curie Actions Individual Fellowship (H2020-MSCA-IF; Global Fellowship ``SATORI'', grant agreement no.~101030927). M.E.T.~and C.G.VdW.~were supported by the U.S.~Department of Energy (DOE; grant no.~DESC0010689). X.Z.~was supported by the National Natural Science Foundation of China (NSFC; grant no.~52172136). This work used resources of the National Energy Research Scientific Computing Center (NERSC; award no.~BES-ERCAP0021021), a DOE Office of Science User Facility supported by the Office of Science of the U.S.~DOE (contract no.~DE-AC02-05CH11231). C.A.B.~thanks Prof.~Jos E.~M.~Haverkort (TU Eindhoven, Netherlands) for useful discussions.


\section*{Data access statement}

The data that support the findings of this study are available from the corresponding author upon reasonable request.



\end{document}